\documentstyle[epsfig]{mn}

\newcommand{\pow}[1]{$^{#1}$}
\newcommand{\ten}[2]{\mbox{${#1}\!\times\!10^{#2}$}}
\newcommand{\sub}[1]{$_{#1}$}
\def\wv{$\lambda$}

\title{A high excitation H\,{\sc ii} region in the faint dwarf elliptical galaxy A0951+68}
\author[R.A. Johnson, A. Lawrence, R. Terlevich and D. Carter]
{R. A. Johnson$^{1,2}$, 
A. Lawrence$^2$,
R. Terlevich$^3$ and
D. Carter$^{3,4}$ \\
$^1$Queen Mary and Westfield College, Mile End Rd., London\\
$^2$Institute for Astronomy, University of Edinburgh, Royal Observatory, Blackford Hill, Edinburgh \\
$^3$Royal Greenwich Observatory, Madingley Road, Cambridge \\
$^4$Astrophysics Group, Liverpool John Moores University, Byrom Street, Liverpool}

\begin{document}

\maketitle

\label{firstpage}

\begin{abstract}
We present the results of BVRI imaging and optical spectroscopy of the
dwarf galaxy A0951+68. The images reveal that, although this galaxy is
classified as a dwarf elliptical, it has some properties that are
similar to dwarf irregular galaxies. It contains two bright knots of
emission, one of which is red and unresolved and the other blue and
resolved. The blue knot also shows a high excitation emission line
spectrum.  The observed line ratios indicate that this is an H\,{\sc
ii} region, although with some line ratios that are border-line with
those in AGN. The emission line luminosity is consistent with
ionisation by a single, very luminous O star, or several smaller O
stars, but the extended blue light in the knot shows that this has
occurred as part of a substantial recent star formation event.  We
find that the metal abundance, while low compared to typical large
galaxies, actually seems to be high for such a low luminosity
dwarf. The position of A0951 in the literature is incorrect and we
provide the correct value.
\end{abstract} 

\section{Introduction}
The galaxy A0951+68 is in the Kraan-Korteweg and Tammann (KKT)
catalogue of nearby galaxies \cite{Kraa:1979,Kraa:1986}, where it is
classified as a dwarf elliptical in the M81 group of galaxies.  From
this catalogue we have selected a complete, distance-limited subset,
with a declination cut-off and excuding irregulars.  A
multi-wavelength survey of this subset, which includes A0951, is in
progress.  During the optical imaging of A0951+68, it was realised
that its position in the KKT catalogue was incorrect. The source of this incorrect position was a paper by Bertola
and Maffei \shortcite{Bert:1974}. Using the photographs presented in
that paper we were able to locate and image A0951+68 at the correct
position, roughly 10 arcminutes from the KKT catalogue position (see
section~\ref{sec:res}). Our images showed two bright knots surrounded
by low surface brightness emission. A search in the literature for
objects at the new position of A0951 revealed two observations prior
to those of the ``discovery'' of Bertola and Maffei.
Karachentseva \shortcite{Kara:1968} discovered the galaxy and named it
Kar61. It was also contained in the list of Mailyan dwarf galaxies
\cite{Mail:1973} where it was called Mailyan 47.

It was no longer clear whether A0951+68 was really faint and nearby,
or more distant and luminous, since the KKT distance was from an HI
measurement at the incorrect postion \cite{Huch:1989}.  HI maps in the
literature do show emission at the correct position but it is unclear
whether this is just a positional coincidence as there is extensive HI
emission in the region around M81.  An independent measurement of
distance was required and to this end we obtained optical spectra of
the two bright knots in the galaxy.  One of the spectra showed very
high-excitation emission lines, from which we were able to confirm
that this galaxy is indeed an M81 group dwarf, but showing interesting
behaviour, atypical of a dwarf elliptical. We have therefore taken the
A0951 distance to be 3.63 Mpc, the same as the M81 distance \cite{Free:1994}.

\section{Observations and data reduction}
\label{sec:obs}
Optical images in the B,V,R and I bands were obtained using the
Jacobus Kapteyn Telescope (JKT) on La Palma in February 1994. The EEV7
CCD used gives a spatial scale of 0.31 arsec/pixel and an image size
of 6\farcm5$\times$ 6\farcm0.  The seeing on this night was poor, FWHM
approximately 2\farcs0.  The spectra were also taken on La Palma, in
January 1995, using the Intermediate Dispersion Spectrograph (IDS) on
the Isaac Newton Telescope (INT). We used a TEK CCD and the R300V
grating which has a dispersion of 3.29 \AA pixel\pow{-1}. The FWHM of
the arc lines is 8.24$\pm$0.62 \AA. The slit width used was 1\farcs5.

The optical images were reduced using the {\sc iraf} software package.
They were debiased and flat-fielded, and calibrated
using Landolt photometric standards, which give Johnson B,V and Kron-Cousins
R and I magnitudes \cite{Land:1983}. Extinction and
colour corrections were determined from these standards.  The spectra
were reduced using the {\sc figaro} software package. Each object and
standard frame was debiased and flatfielded, and the frames were
cleaned of cosmic rays. The sky was removed from each frame by
subtracting a polynomial fitted to the object free parts of each
frame. 1D spectra were extracted from each frame and wavelength
calibrated. The wavelength calibration was provided by Cu-Ar and Cu-Ne
lamps which were observed once either side of the observations. The
wavelength values we use are for standard air conditions. The
wavelength scale shifted by $\approx$ 1\AA\ between the two lamp
spectra. The average of the two wavelength scales was used to
calibrate the object and standard frames that were observed between
the two lamp spectra.  The maximum error on the wavelength calibration
is therefore $\approx$ 0.5\AA.  The standard and object frames were
corrected for atmospheric extinction
using the standard extinction curve for La Palma \cite{King:1985} and the measurements for
that night from the Carlsberg Automatic Meridian Circle.
The flux standards used for the spectroscopic observations were SA 29
130 \cite{Oke:1974} and Feige 34 \cite{Ston:1977}.

\section{Results}
\label{sec:res}
\subsection{Morphology, position and colours}
Figure 1 shows the R image of A0951. There are two
bright knots of emission, A0951A and A0951B, superimposed on diffuse
low surface brightness emission. The knot A0951A appears to be close
to the peak of the diffuse emission, whereas A0951B is towards the
edge of the galaxy. Figure~\ref{fig:cuts} shows cuts through the
galaxy and both bright knots in the sky subtracted R and B images.
This also shows the central position of A0951A and highlights the
difference in colour between the two knots.
\setcounter{figure}{1}
\begin{figure*}
\vbox to 11cm{\vfil
  \epsfxsize 8cm
   \epsfbox{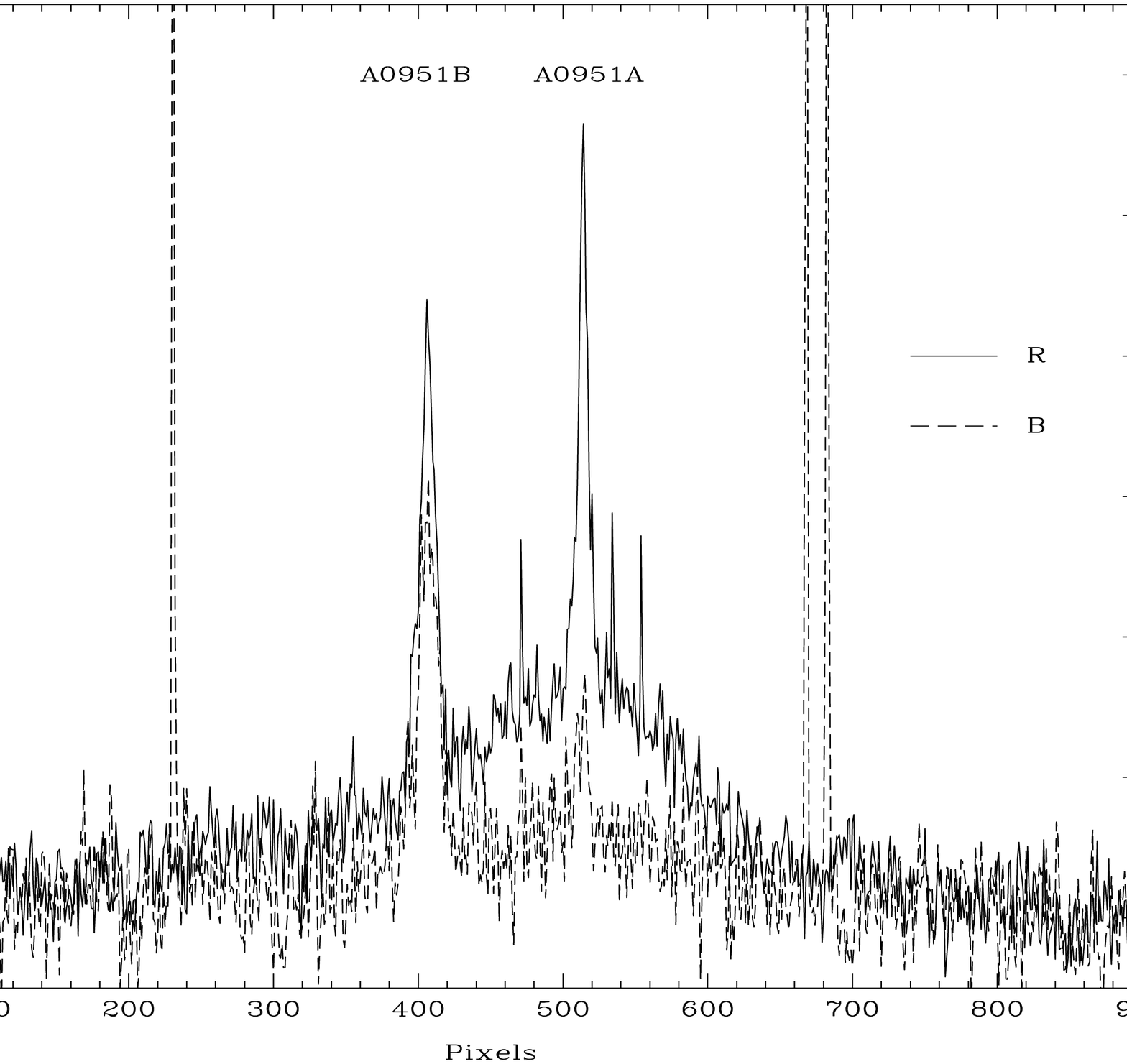}
  \caption{Cuts through the galaxy in the B and R images}
  \label{fig:cuts}
  \vfil}
\end{figure*}
To see whether the two knots of emission were resolved we fitted
Gaussian profiles to each knot and to three stars in each frame.
Table~\ref{tab:res?} gives the average value of sigma in arcseconds
for the stars and for the two knots.  The brackets indicate an
estimation since there was not enough signal-to-noise to enable a
gaussian fit.  A0951A is unresolved at all wavelengths, A0951B is
clearly resolved in B,V and R.
\begin{table}
  \caption{Comparison of $\sigma$ (arcsec) of Gaussian fits to stars and knots}
  \label{tab:res?}
\begin{tabular}{||l|l|l|l||} 
   & $\sigma_{\small \rm stars}$ & $\sigma_{\small \rm A0951A}$ & $\sigma_{\small \rm A0951B}$ \\ \hline
 B & 0.98  & (1.1)    & 1.38 \\
 V & 1.02  & 0.94   & 1.30 \\
 R & 0.78  & 0.79   & 1.20 \\
 I & 0.69  & 0.76   & (0.8)    \\ \hline
\end{tabular}
\end{table}

Table~\ref{tab:mags} gives the magnitudes of the knots and the diffuse
emission.  To measure the knot magnitudes the images were first
smoothed so that the p.s.f in each image was the same as that in the
image with the worst seeing.  The magnitudes for the knots were
measured inside a circular aperture that enclosed all of the knot
emission in all filters.  The aperture radii for knot A and knot B
were 3\farcs25 and 4\farcs03 respectively. The emission underlying
each knot was found by taking the mean in an annulus surrounding the
aperture and this underlying emission was subtracted from the knot
magnitudes.  The magnitude of the diffuse emission was found down to
the surface brightness contour where the galaxy emission became
indistinguishable from the sky. The surface brightness of the contour
enclosing the diffuse emission is given in the table, as well as the
approximate size of the galaxy down to this surface brightness. The
knot magnitudes have been subtracted from the diffuse magnitude.

Table~\ref{tab:cols} gives the colours of the knots and the diffuse
emission. To find the colours of the diffuse emission, its magnitude
in the smoothed images was found, within a fixed  area that contained 
significant diffuse emission in all the images.

\begin{table*}
\caption{Magnitudes}
\label{tab:mags}
\vbox to4cm{\vfil
\begin{tabular}{||l|l|l|l|l|l||}
filter & Knot A & Knot B & Diffuse & limiting contour  & approximate \\ 
       &        &        &         & mag arcsec\pow{-2}& size (arcsec)\\\hline 
B  & 21.63 & 19.81 & 17.22 & 26.5 & 43$\times$30\\
V  & 20.67 & 19.80 & 15.68 & 26.5 & 89$\times$60\\
R  & 20.06 & 19.69 & 15.10 & 26   & 94$\times$63\\
I  & 19.77 & 20.04 & 14.64 & 26   & 92$\times$56\\ \hline
\end{tabular}
\vfil}
\end{table*}

\begin{table}
\caption{Colours}
\label{tab:cols}
\begin{tabular}{||l|l|l|l||}
colour & Knot A & Knot B & Diffuse \\ \hline
B-V & 0.96 & 0.01 & 0.53 \\
V-R & 0.61 & 0.11 & 0.49 \\
V-I & 0.91 & -0.25 & 0.99 \\ \hline
\end{tabular}
\end{table}

The digitized POSS plates were used to find the RA and dec of bright
stars in the optical image and from these we calculated the positions
of the bright knots A0951A and A0951B to be
$09^{\rm{h}}~53^{\rm{m}}~00\fs01$, $+68^\circ~49'~51\farcs3$ and
$09^{\rm{h}}~53^{\rm{m}}~04\fs86$, $+68^\circ~50'~10\farcs5$ (B1950)
respectively.  This is $\approx$ 10 arcmin away from the position
quoted by Bertola and Maffei \shortcite{Bert:1974}.

\subsection{Spectrum}
\begin{figure*}
  \vbox to 11.5cm{\vfil
  \epsfxsize 8cm
  \centerline{\epsfbox{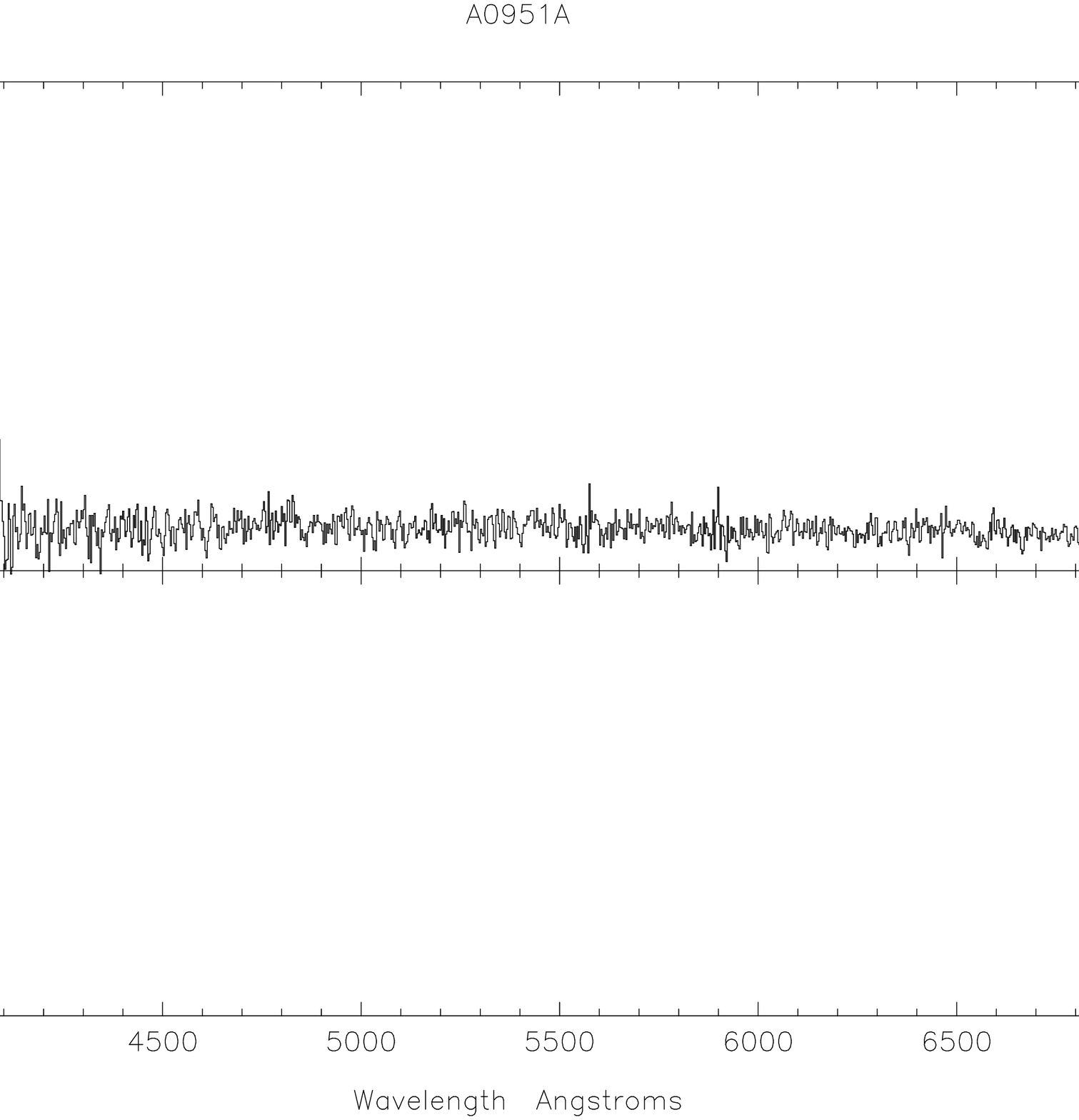}}
  \caption{Spectrum of A0951A}
  \label{fig:nasp}
  \vfil}
\end{figure*}

\begin{figure*}
  \vbox to 11.5cm{\vfil
  \epsfxsize 8cm
  \centerline{\epsfbox{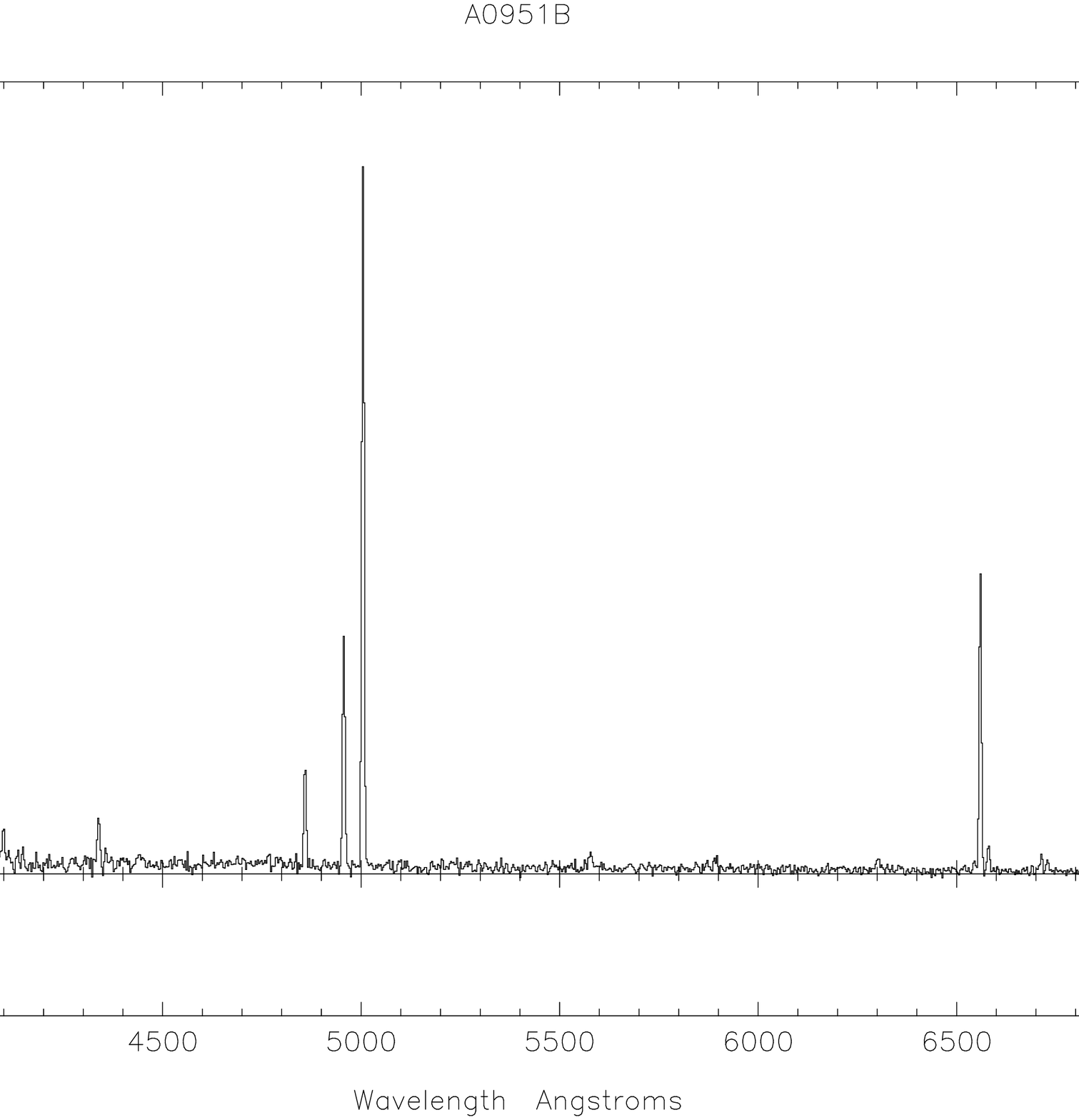}}
  \caption{Spectrum of A0951B}
  \label{fig:nbsp}
  \vfil}
\end{figure*}

Figures~\ref{fig:nasp} and~\ref{fig:nbsp} show the spectra of the two
bright knots.  Both spectra have featureless continua, but A0951B
contains strong high excitation emission lines.  The observed value of
3.19 for the Balmer decrement in A0951B is greater than the Case B
recombination value of 2.86 (for T\sub{e}=10000K and N\sub{e}=100
cm\pow{-3}). Using the reddening law in Savage and Mathis
\shortcite{Sava:1979} this implies an E(B-V) of 0.1. The galactic
reddening in this direction gives an E(B-V) of 0.04 \cite{Burs:1984}.
Table~\ref{tab:flxs} lists the line identification, the observed
wavelengths and the observed and de-reddened (assuming E(B-V)=0.1)
line fluxes in A0951B.
\begin{table*}
  \vbox to 9cm{\vfil
  \epsfxsize 11cm
\caption{A0951B Line Fluxes}
\label{tab:flxs}
\begin{tabular}{||llll||}
Line ID & $\lambda$\sub{\small obs} & F($\lambda$)/F(H$\beta$)\sub{\small obs} &F\sub{r}($\lambda$)/F\sub{r}(H$\beta$) \\ \hline
 H$\delta$ 4102             & 4099.2 & 0.40 & 0.42              \\
 H$\gamma$ 4340             & 4338.9 & 0.47 & 0.50              \\
 H$\beta$  4861             & 4858.9 & 1.0  & 1.0               \\
 {[}\hbox{O\,{\sc iii}}] 4959 & 4956.4 & 2.72 & 2.70              \\
 {[}\hbox{O\,{\sc iii}}] 5007 & 5004.5 & 7.75 & 7.68              \\
 {[}\hbox{N\,{\sc ii}}]  6548 & 6545.9 & 0.11 & 0.10              \\
 H$\alpha$ 6563             & 6559.7 & 3.19 & 2.86              \\
 {[}\hbox{N\,{\sc ii}}] 6584  & 6580.1 & 0.30 & 0.27              \\    
 {[}\hbox{S\,{\sc ii}}] 6717  & 6713.1 & 0.14 & 0.12              \\
 {[}\hbox{S\,{\sc ii}}] 6731  & 6728.4 & 0.09 & 0.08              \\ 
\multicolumn{4}{||l||}{Upper limits on undetected lines} \\ 
 {[}\hbox{O\,{\sc ii}}] 3727  &        & $\le$ 1.38 & $\le$ 1.52   \\
 {[}\hbox{O\,{\sc iii}}] 4363 &        & $\le$ 0.17 & $\le$ 0.18   \\
 \hbox{He\,{\sc ii}} 4686   &        & $\le$ 0.14 & $\le$ 0.14   \\
 {[}\hbox{O\,{\sc i}}I] 6300  &        & $\le$ 0.11 & $\le$ 0.10   \\
\multicolumn{4}{||l||}{Observed H$\beta$ flux = \ten{2.63}{-15} erg s\pow{-1}cm\pow{-2}} \\
\multicolumn{4}{||l||}{Reddening corrected H$\beta$ flux = \ten{3.71}{-15} erg s\pow{-1}cm\pow{-2}} \\ \hline
\multicolumn{4}{||l||}{Col 1 Line Identification, Col 2 observed wavelength} \\
\multicolumn{4}{||l||}{Cols 3 \& 4 Observed and reddening corrected line fluxes relative to H$\beta$=1} \\
\end{tabular}  
\vfil}
\end{table*}

To provide checks of the spectrum and image measurements, the flux of
the spectra in wavelength bands roughly corresponding to the Johnson
B,V and Kron-Cousins R,I filters was calculated and converted to
magnitudes.  Taking into account that the area of the object covered
by the slit is less than the aperture used for the photometry, these
magnitudes are consistent with those calculated from the images.

The strongest lines in the spectrum, H$_{\beta}$, [\hbox{O\,{\sc
iii}}]\wv\wv4959,\linebreak 5007, and H$_{\alpha}$ were used to calculate the
recession velocity of A0951. Applying a correction for the earth's
orbit gives a heliocentric velocity of -134.73$\pm$3.29
kms\pow{-1}$\pm$ 30kms\pow{-1}. The first error here is the root mean
square fitting error on the calculated velocity, the second is a
systematic error due to the shift in the wavelength calibration during
the night mentioned in section~\ref{sec:obs}. 

The measured FWHM of the H$\alpha$ line is 9.02\AA. Recalling that the FWHM of
the arc lines is 8.24$\pm$0.62 \AA, this implies that the H$\alpha$ line
is unresolved, and that, at 1$\sigma$, the intrinsic
FWHM of H$\alpha$ $<$ 272 kms\pow{-1}. This is consistent with values
of $<$ 120 kms\pow{-1} found in other H\,{\sc ii} regions.

\section{Discussion}
A0951+68 has previously been classified as a dwarf elliptical from its
smooth and symmetric appearance on photographic plates. The presence
of A0951B was not known and the nature of A0951A was unclear -
Karachentseva et al. \shortcite{Kara:1968} thought it was a foreground
star, whilst Bertola \& Maffei \shortcite{Bert:1974} were unable to
tell whether it was a star or the nucleus of the galaxy.  The knot
A0951A is close to the centre of the inner isophotes and this part of
the galaxy has the appearance of a nucleated dwarf
elliptical. However, A0951B is a high excitation H\,{\sc ii} region
(see section~\ref{sec:a0951b}), and this and the outer isophotes look
more like an Im type irregular galaxy.  In their atlas of Virgo
galaxies, Sandage and Bingelli \shortcite{Sand:1984} note that there
are a few cases of unsure classification between dE and Im galaxies,
but are unable to say whether this is due to a real evolutionary link
between the two types. The absolute V magnitude of A0951 within $\mu V
= 26.5$ is -12.12.  This is at the faint end of a sample of M81 dwarf
irregulars observed by Miller and Hodge \shortcite{Mill:1994}, which
have M\sub{\rm V} between -11.9 and -16.7.

Another property of A0951 that is similar to dwarf irregulars is its
probable HI content. Several groups have published maps of the neutral
hydrogen emission in the vicinity of M81 which include the position of
A0951 \cite{Huls:1979,Appl:1981,Yun:1994}. These all show neutral
hydrogen at the position of A0951. The paper by van der Hulst
\shortcite{Huls:1979} contains velocity channel maps which show HI at
the position of A0951 with a velocity of -140 kms\pow{-1}, consistent
with the velocity measured from our optical spectrum. However the
large amount of HI in this region and the low spatial resolution of
the maps means we cannot be entirely sure that the gas is coincident
with A0951.

\subsection{Diffuse emission}
\label{sec:diff}
\begin{figure*}
\vbox to 15cm{\vfil
  \epsfxsize 10cm
  \epsfbox{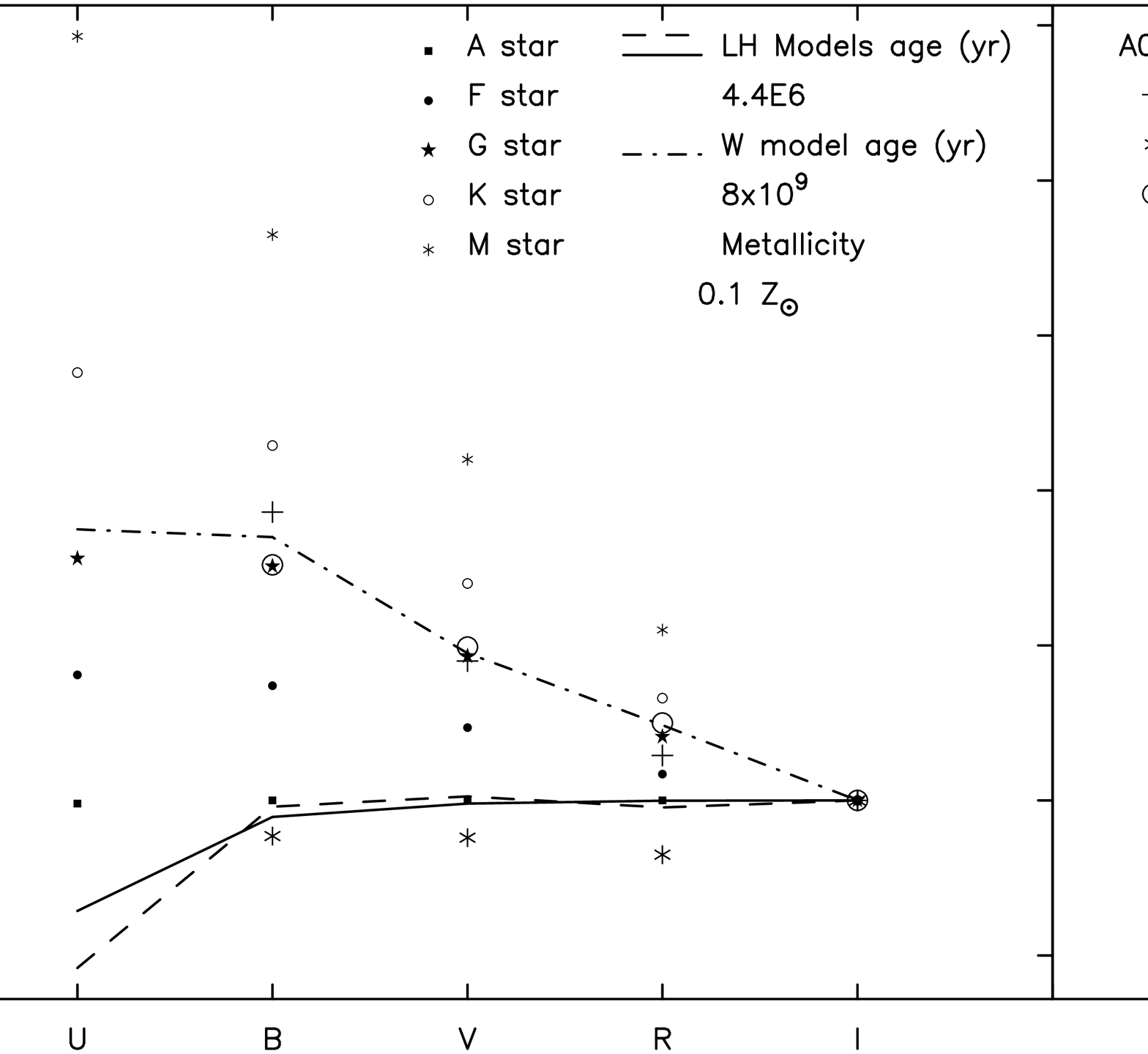}
   \caption{Colour of A0951 knots and diffuse emission. The ordinate is the 
colour relative to I, i.\ e.\ at B the value is B-I. The stellar population 
models are those in Worthey (1994) and Leitherer \protect\& Heckman (1995) 
(see sections 4.1 and 4.3). Both models are low metallicity, 0.1Z\sun}
  \label{fig:a0951cols}
  \vfil}
\end{figure*}

The colours of A0951 can be used as well as the morphology to pin down
its type.  Figure~\ref{fig:a0951cols} shows the colours of the two
knots and the diffuse emission. The ordinate gives the colour relative
to I. The positions of main sequence stars are
shown, as are models of old and young stellar populations with a
metallicity (0.1Z\sun) consistent with that derived from the emission line
spectrum of knot B (see section~\ref{sec:a0951b}).  The 8Gyr model is
from Worthey \shortcite{Wort:1994} and fits well the colours of the
diffuse emission and of knot A. A slightly better fit is provided
assuming a metallicity of 0.03Z\sun, but this is lower than found from
the emission lines. The colours of the diffuse emission are very
similar to those of a G star.  The B-V colour of 0.53 of the diffuse
emission is bluer than that of dwarf ellipticals in the Virgo and
Fornax clusters \cite{Cald:1983,Cald:1987} which have a mean B-V of
0.78.  Gallagher and Hunter \shortcite{Gall:1987} present
colour-colour diagrams in BVRI for a sample of irregular and amorphous
galaxies, and A0951 has the same postion on these diagrams as these
galaxies.  Although these galaxies are more luminous than A0951, the
colours of dwarf irregulars are similar as indicated by the mean B-V
colour of 0.47 of dwarf irregulars in Virgo \cite{Both:1986}, and a
mean B-V of 0.37 in a sample of dwarf irregulars also observed by
Hunter and Gallagher \shortcite{Hunt:1985}.

\subsection{A0951A}
Knot A has much redder colours than Knot B. The V-I colour is similar
to that of the diffuse emission; the other colours are somewhat
redder. As we do not have a redshift for this knot there is the
possibility that it is a superimposed foreground star - in
Figure~\ref{fig:a0951cols} it can be seen that the colours of A0951A
are closest to those of a G star. The B-V colour of A0951A is at the
far red end of the colours of dwarf ellipticals in the Virgo and
Fornax clusters. The B-V, V-R and R-I colours of A0951A are similar to
those in a sample of low luminosity early type galaxies observed by
Prugniel et al \shortcite{Prug:1993}. Some dwarf ellipticals are
nucleated and these nuclei have similar colours to the surrounding
emission. The colours and morphology of A0951A are reminiscent of
these dwarf elliptical nuclei.

The spectrum of A0951A is flat and featureless. Held and Mould
\shortcite{Held:1994} show spectra of dwarf elliptical nuclei. On the
whole they are not as flat as A0951A. The most prominent feature in
these spectra is the 4000\AA\ break, unfortunately at this wavelength
the spectrum of A0951A has very low signal to noise and although
there is a hint of a break it is impossible to say whether it is
real.  The signal to noise in our spectrum is too low to observe the
absorption features seen in these spectra.

\subsection{A0951B}
\label{sec:a0951b}
The spectrum of A0951B is very striking. The value of the
[\hbox{O\,{\sc iii}}]\wv5007/H$\beta$\wv4861 ratio is as high as that seen
in active galaxies, but the [\hbox{N\,{\sc ii}}]\wv6584/H$\alpha$6563
ratio is much lower, suggesting low metallicity.  The different
types of emission line object, H\,{\sc ii} regions and galaxies,
active galaxies and planetary nebulae can be separated by their
positions on emission line ratio diagrams such as those of Veilleux
and Osterbrock \shortcite{Veil:1987} and Baldwin, Phillips and
Terlevich \shortcite{Bald:1981}. In these diagrams A0951 occupies the
area populated by H\,{\sc ii} regions and H\,{\sc ii} galaxies, though
the [\hbox{O\,{\sc iii}}]\wv5007/H$\beta$\wv4861 ratio in this galaxy
is amongst the highest seen. A0951B is somewhat peculiar in that the
few other H\,{\sc ii} galaxies and regions that have a similar
[\hbox{O\,{\sc iii}}]\wv5007/H$\beta$\wv4861 ratio have even lower
[\hbox{N\,{\sc ii}}]\wv6584/H$\alpha$6563 ratios. The value of
[\hbox{N\,{\sc ii}}]\wv6584/H$\alpha$6563 in A0951B puts it in the
transition region between H\,{\sc ii} regions and AGN. Also (see
below), although its metal content is low compared to typical large
galaxies, it is actually high compared to other H\,{\sc ii} galaxies
of the same luminosity.

The observed emission line ratios in H\,{\sc ii} regions depend upon the 
properties of the ionized gas and on the ionizing stars. Some of these
properties can be calculated directly from the observations, others can
be found by comparing the observations with models of H\,{\sc ii} regions.

From the observed emission lines in A0951B we can deduce the electron
temperature and density of the gas, and the number of ionizing
photons.  The electron temperature in the O\pow{++} region can be
found from the ratio of the [\hbox{O\,{\sc iii}}]$\lambda\lambda$4959,
5007 doublet and [\hbox{O\,{\sc iii}}]$\lambda$4363. The fluxes in
table~\ref{tab:flxs} give an upper limit to the temperature of 15500
K. The electron density can be estimated from the [\hbox{S\,{\sc ii}}]
$\lambda\lambda$6717, 6731 doublet; however this loses sensitivity at
typical H\,{\sc ii} galaxy densities of less than $\approx$ 200
cm\pow{-3}. The density in A0951B is in this low density limit.  The
number of hydrogen ionizing photons, Q(H), emitted by the ionizing
stars is given by the relationship ${\rm
Q(H)(s^{-1})}=7.31\times10^{11}{\rm L_{H\alpha}(ergs^{-1})}$
\cite{Oste:1989}. Using the reddening corrected H$\alpha$ luminosity
for A0951B gives ${\rm Q(H)}=1.20\times10^{49}$ s\pow{-1}.  This
number of ionizing photons could be produced by one star of spectral
type O7 (T\sub{\rm e}=45200 K) or several stars of later type.

Two of the most recent H\,{\sc ii} region models are those of Cervino
and Mas-Hesse \shortcite{Cerv:1994} and Garc\'{i}a-Vargas, Bressan and
D\'{i}az \shortcite{Garc:1995}.  Both these models vary the ionization
parameter, the effective temperature of the ionizing cluster
(T\sub{\rm e}), and the metallicity of the gas, and calculate the
emission line ratios produced.  The ionization parameter, U, is
defined as the ratio between the density of the ionizing photons and
the density of particles ($U=Q(H)/4\pi cn_{H}r^{2}$), where {\it n}\sub{\rm H} is
the density of ionized hydrogen).  This depends on the number of
ionizing photons, the density of hydrogen atoms and the radius of the
H\,{\sc ii} region, so knowing the first two of these parameters would
enable us to estimate the radius of the H\,{\sc ii} region. In fact we
only have an upper limit on the density, but the interesting thing to
work out is how the radius of the H\,{\sc ii} region compares to the
Stromgren radius, and the ratio of these two radii only depends upon
(density)\pow{1/6}.

The aim of Cervino and Mas-Hesse was to calculate the dependence of
the most commonly observed emission lines, those of oxygen, on these
parameters.  From our limits on the oxygen emission line ratios in
A0951 we can deduce from their models that T\sub{\rm e} is 
between 45000 and 55000 K and the oxygen abundance, log(O/H), between
-4.2 (1/12 solar) and -3.8 (1/5 solar). They do not say what value of
ionization parameter they use.  They show the parameters derived from
their models for a sample of H\,{\sc ii} galaxies \cite{Terl:1991}. Comparison
of A0951B with these galaxies shows that it has a similar metal
content to the average H\,{\sc ii} galaxy, but a much lower luminosity.

Garc\'{i}a-Vargas et al.\ model the behaviour of more emission lines,
and look at the line ratios for different ages of the ionizing star
cluster (and hence different T\sub{\rm e}) and for varying metallicity and
ionization parameter.  Their models suggest that A0951 has a
metallicity, Z, between 0.001 (1/17 Z\sun) and 0.008 (1/2 Z\sun), an
age between 2 and 5.2 Myrs (implying a T\sub{\rm e} between 40000 and
50000K), and an ionization parameter, log U, between -2.7 and -2.4.
From this we can deduce that the radius of the H\,{\sc ii} region is the same
as the radius of a Str\"{o}mgren sphere of an O7 star. The range of
effective temperatures implied by these models, combined with the
number of ionizing photons is consistent with ionization by a single 
such O star, or a few, later-type O stars.

It is interesting to compare A0951 with GR8, a dwarf irregular in the
local group \cite{Skil:1988}. While their total absolute magnitudes
(-12.1 and -10.7) and H\,{\sc ii} region luminosities (L(H$\beta$)
\ten{1.4}{36} and \ten{5.8}{36}) are comparable, the metal content of
A0951 seems to be much higher than that of GR8 or other star forming
regions of the same total luminosity. This may suggest a different chemical
evolution history for A0951, perhaps related to the fact that A0951 is
not a dwarf irregular.  Clearly a better estimate of the metal
abundances is needed.
 
The H$\alpha$ luminosity of A0951B is {\rm $1.64\times10^{37}$ ergs
s\pow{-1}}.  This is at the low end of the range of luminosities seen
in dwarf irregulars.  It is interesting to note that one of the dwarf
irregular H\,{\sc ii} reigons observed by Hunter and Gallagher
\shortcite{Hunt:1985} has an ionizing source of a single O star, and
similar line ratios to A0951B with [\hbox{O\,{\sc
iii}}]\wv5007/H$\beta$\wv4861 = 6.18.  The H$\beta$ equivalent width
in A0951B is 92.8 \AA\ which is higher than the mean equivalent width
in the Terlevich et al. \shortcite{Terl:1991} sample of H\,{\sc ii}
galaxies and is indicative of a young stellar population.

The emission-line region that we have been discussing occurs somewhere
inside the resolved blue knot A0951B, which is clearly much bigger and
more luminous than a single O star.  The colours of knot A0951B are
similar to those in resolved regions of star formation found in some
Virgo dwarf irregulars by Bothun et al. \shortcite{Both:1986}. A
recent paper by Leitherer \& Heckman \shortcite{Leit:1995} shows the
change in colours of a model starburst galaxy with metallicity and
age.  Their model for a metallicity of 0.1 Z\sun\ and an age of 4.4
Myr (chosen to be consistent with the values deduced from the
spectrum) is shown in Figure~\ref{fig:a0951cols}. The solid line is
for an instantaneous burst and the dashed line for continuous star
formation. It can be seen that the model colours are reasonably
similiar to the colours of A0951B.  We conclude that A0951B is a
region of very recent star formation which in particular contains an
H\,{\sc ii} region ionized by at most a few O stars, with an effective
temperature of around 45000-50000K.

\section{Conclusions}
Optical images and spectroscopy of the M81 dwarf A0951+68 have
provided a few surprises, not least that the published position of the
galaxy was incorrect. We have calculated a new position. Another
suprise was the presence in this supposed dwarf elliptical galaxy of
an H\,{\sc ii} region, with very high excitation.  H\,{\sc ii} region
models suggest that the ionizing source is a single very luminous O7
star, or a few later-type O stars. The colours of the knot A0951B,
which contains this H\,{\sc ii} region, are consistent with it being a
star formation region.  The colours of the diffuse emission in A0951,
the presence of the star forming region and also the fact that it
probably contains H\,{\sc i} gas make A0951 more like a dwarf irregular
than a dwarf elliptical.  However, the knot of emission at the centre
of the inner isophotes (A0951A) has colours that are similiar to those
of low luminosity ellipticals and is reminiscent of the nuclei seen in
some dwarf elliptical galaxies.

\section*{Acknowledgements}
The Jacobus Kapteyn and Isaac Newton telescopes are operated on the island
of La Palma by the Royal Greenwich Observatory in the Spanish Observatorio
del Roque de los Muchachos of the Institute de Astrofisica de Canarias.
The {\sc iraf} and {\sc figaro} software packages used to reduce the data 
are distributed by NOAO and Starlink respectively. RAJ acknowledges
receipt of a PPARC studentship.

\end{document}